\documentclass[12pt,a4]{article}
\usepackage{latexsym}

\newcommand{\bea}{\begin{eqnarray}}
\newcommand{\eea}{\end{eqnarray}}

\title{Normal Modes, Quasi-normal Modes and Super-radiant Modes 
for Scalar Fields    
in Kerr anti-de Sitter Spacetime }
\author{Masakatsu KENMOKU\thanks{kenmoku@asuka.phys.nara-wu.ac.jp} 
\ \\
Department of Physics, 
Nara Women's University, Nara 630-8506, Japan
}
\date{\empty}
\begin{document}
\maketitle
\abstract{
Normal modes, quasi-normal modes and super-radiant modes 
are studied to clarify the total dynamics 
for complex scalar fields in Kerr anti-de Sitter black hole spacetime.  
Orthonormal relations and quasi-orthonormal relations are obtained 
for normal modes and quasi-normal modes.  
Mode expansions are done and the conserved quantities are studied.  
Any modes are shown to be separated into two groups, 
physical modes and unphysical modes, by the zero mode line. 
Zero modes themselves do not exist as normalizable modes 
with the correct boundary condition.  
The allowed physical modes exclude the 
super-radiant instability modes in rotating black hole spacetime. 
The result is consistent with the co-rotating frame consideration. 
}

\section{Introduction}
\renewcommand{\theequation}{\thesection.\arabic{equation}}
\setcounter{equation}{0}
Black holes have many interesting features in 
theoretical as well as in observational investigations. 
As the development of observation techniques and devices, 
many candidates of black holes have been observed including 
super-massive black holes in the center of galaxies 
\cite{eckart001,herrnstein001,MBH003}. 
These black holes are expected to be described well 
by the exact axisymmetric solutions of Einstein's equation 
\cite{exact001}.    
Schwarzschild, Kerr and Reissner-N\"{o}rdstrom solutions are known as  
massive, rotating and charged black holes in 
(3+1)-dimensional spacetime. 
Incorporating the cosmological constant to them, 
Kerr de Sitter and Kerr anti-de Sitter (Kerr-AdS) solutions 
are also known, which 
are interesting in views of the recent observation 
of cosmological term in WMAP \cite{WMAP001} 
and AdS/CFT correspondence \cite{maldacena001}. 
Higher dimensional multi-rotating black hole solutions 
\cite{myers001,gibbons001,chen001}
are also interesting in views of recent development in 
string theory \cite{green001,polchinski001}, 
brane world \cite{randall001} and M-theory 
\cite{kaku001}.  
 
The interaction of matter fields with black holes are important 
in observational and theoretical understanding of black holes. 
Perturbation of matter fields in the black hole spacetime 
is investigated substantially 
\cite{birrell001,chandra001,kokkotas001}. 
The massless field equations for the field of 
scalar, Dirac, Maxwell, Rarita-Schwinger and tensor  
are summarized as Teukolsky equations \cite{teukolsky000}
and studied extensively  
\cite{leaver001,takasugi001} 
\footnote{Teukolsky equation is categorized as Heun's equation in
Mathematics \cite{heun001}.}.     
One of important investigations 
is the stability problem of black holes. 
Perturbative stability has been studied 
in (3+1) and higher dimensional spacetime 
\cite{kodama001}.
The present status of the stability issue for (3+1)-dimensional 
black holes is that the black holes of 
Schwarzschild, Kerr, Reisner-N\"{o}rdstrom are stable 
\footnote{
The instability of scalar fields in rotating spacetime 
without cosmological constant is reported  
\cite{detweiler001,nambu001,dotti001}.}. 
As to Kerr-AdS black hole spacetime, 
large rotating black holes are studied to be stable 
\cite{hawking000} but small black holes  
are unstable under the condition giving 
rise to super-radiant instability in 
the scattering of scalar fields with black holes 
\cite{teukolsky001,cardoso001,cardoso002}. 
Asymptotic AdS spacetime 
plays the role of 
a natural reflecting mirror which amplifies the scattered wave 
repeatedly to lead the instability, which is called as  
black hole bomb \cite{press001}.   
For the incident wave with the frequency $\omega$ and the azimuthal 
angular momentum $m$ upon a target rotating black hole,  
the scattered wave is amplified 
if the super-radiant instability condition is satisfied:  
\begin{eqnarray}
{\rm Re}\,\omega-\Omega_{\rm H}m<0\ \ \ \mbox{\rm with}\ \ \  
0<{\rm Re}\,\omega\ ,  
\end{eqnarray}
where $\Omega_{\rm H}$ denotes  
the angular velocity of black holes at horizon (see Eq.(\ref{velocity001})). 
In higher dimensions, 
the situation is similar to (3+1)-dimensional spacetime so that  
rapidly rotating Kerr-AdS black holes  
are reported to be unstable \cite{kunduri001}. 

The black hole thermodynamics 
\cite{bekenstein001,bardeen001,hawking001,carlip001} 
is also an important issue 
in the interaction of matter fields with black holes. 
The microscopic understanding of the black hole thermodynamics 
has been studied extensively 
in string theory \cite{strominger001}, 
conformal field theory and 
brick wall model \cite{thooft001}. 
The brick wall model is the model to 
built the brick wall at the horizon in order to confine the scalar fields 
 around the black hole and 
sum up all the eigenstates to calculate the entropy 
according to the standard statistical mechanics. 
In the brick wall model, 
the problem is that the Boltzmann weight cannot 
be well-defined if the super-radiant instability occurs 
in rotating black hole spacetime in (3+1)-dimensions 
\cite{mukohyama001,mukohyama002} 
and in (2+1)-dimensional BTZ \cite{btz001}
black holes \cite{ichinose001,swkim001,fatibene001,ho001}. 

In view of these severe super-radiant instability problem, 
the purpose of this paper is to make clear the dynamics of 
scalar fields in rotating black hole spacetime 
as exact as possible in analytical method.   
We have already studied eigenvalue problem of normal modes   
for scalar fields analytically and numerically in (2+1)-dimensional 
BTZ black hole spacetime \cite{kenmoku003,kuwata001} to show that 
the super-radiant instability does not occur, which 
is consistent with the negative imaginary part of the quasi-normal 
frequency by Birmingham \cite{birmingham001}. 
We extend our previous method to (3+1)-dimensional 
rotating black hole spacetime, especially to Kerr-AdS black holes.  
Our strategy is to study the total dynamics of scalar fields 
with the Dirichlet or the Neumann boundary condition at horizon 
for normal modes (see Eq.(\ref{normalboundary001})) and 
with the ingoing (into black hole) boundary condition at horizon 
for quasi-normal modes (see Eq.(\ref{quasinormalboundary001})), 
in order to clarify the independence and the completeness of 
each modes.  
Any modes are separated into two groups, physical modes and 
unphysical modes, by the zero mode line defined as  
\begin{eqnarray}
0={\rm Re}\,\omega-\Omega_{\rm H}m\ .  
\end{eqnarray}
Zero modes themselves do not exist as normalizable modes with 
the correct boundary condtion.  
By non-existence of zero modes and 
assumed analyticity of rotation parameter, 
the allowed physical mode region is derived:
\begin{eqnarray}
0<{\rm Re}\,\omega-\Omega_{\rm H}m \ , 
\end{eqnarray}
where the rotation parameter is defined for the real value of the horizon  
$0<r_{-}<r_{+}$ (see Eq.(\ref{horizon001})). 
The allowed physical modes 
exclude the super-radiant instability modes  
in rotating black hole spacetime.  

The organization of this paper is the following. 
In section 2, Kerr-AdS spacetime and 
equations of scalar fields will be  
reviewed and summarized for the following convenience. 
In section 3, orthonormal relations and completeness relations 
of normal modes will be studied. 
Normal mode expansion, quantization and the conserved quantities will 
also be studied. 
In section 4, quasi-orthonormal relations for quasi-normal 
modes will be studied in the similar method for normal modes. 
The fluxes of energy and angular momentum are also studied. 
In section 5, zero modes and super-radiant modes will be studied in
detail. This section is one of the most important parts of this paper.  
Final section is to summarize the results. 
Rotating black holes in co-rotating frame will be studied in appendix.   

\section{Kerr-AdS spacetime and equations of scalar fields}
\setcounter{equation}{0}

In this section, 
Kerr-AdS spacetime and equations of scalar fields are reviewed and summarized for the
following convenience. Throughout this paper, the natural unit 
is used: $c=\hbar=G=1$.     

\subsection{Kerr-AdS spacetime}
The Einstein-Hilbert action with negative cosmological constant 
$\Lambda$ is 
\begin{eqnarray}
I_{G}=\frac{1}{16\pi }\int d^4x \sqrt{-g}\, (R-2\Lambda)\ .
\end{eqnarray}
The vacuum Einstein's equations for this action are 
\begin{eqnarray}
R_{\mu\nu}-\frac{g_{\mu\nu}}{2}R+{g_{\mu\nu}}{\Lambda}=0\ .
\end{eqnarray}
The (3+1)-dimensional Kerr-AdS metric is given by Carter  
\footnote{The Carter metric is related to the Boyer-Lindquist metric  
\cite{boyer001}:  
 $t_{\rm BL}=\Xi\, t_{\rm Carter}$ and 
$\omega_{\rm BL}=\omega_{\rm Carter}/\Xi$. } ,
\begin{eqnarray}
ds^2=
&-&\frac{\Delta_{r}}{\rho^2}
\left(dt-\frac{a \sin^2{\theta}}{\Xi}d\phi\right)^2
+\frac{\Delta_{\theta}\sin^2{\theta}}{\rho^2}
\left(adt-\frac{r^2+a^2}{\Xi}d\phi\right)^2\nonumber\\
&+&\frac{\rho^2}{\Delta_{r}}dr^2+\frac{\rho^2}{\Delta_{\theta}}d\theta^2
\ , \label{line-element}
\end{eqnarray}
where 
\begin{eqnarray}
&&\Delta_{r}=(r^2+a^2)(1+r^2\ell^{-2})-2Mr \ , \ 
\Delta_{\theta}=1-a^2\ell^{-2}\cos^2{\theta}\ , \nonumber \\
&&\rho^2=r^2+a^2\cos^2{\theta}\ \ \ , \ \ \  
\Xi=1-a^2\ell^{-2}\ ,\label{metric}
\end{eqnarray}
with $\ell=\sqrt{-3/{\Lambda}}$ denotes the cosmological
parameter and $a=J/M$ demotes the rotation 
parameter per unit black hole mass $M$.    
The metrics in the standard form,    
$
ds^2=g_{tt}dt^2+g_{\phi \phi}d\phi^2+2g_{t\phi}dtd\phi+g_{rr}dr^2 
+g_{\theta\theta}d\theta^2,
$
is given by 
\begin{eqnarray}
g_{tt}&=&
\frac{1}{\rho^2}(-\Delta_{r}+a^2\sin^2{\theta}\Delta_{\theta})\ , \ 
g_{t\phi}=\frac{a\sin^2{\theta}}{\rho^2\Xi}
(\Delta_{r}-(r^2+a^2)\Delta_{\theta})\ , \nonumber\\ 
g_{\phi\phi}&=&
\frac{\sin^2{\theta}}{\rho^2\Xi^2}
(-a^2\sin^2{\theta}\Delta_{r}+(r^2+a^2)^2\Delta_{\theta})\ , \ 
g_{rr}=\frac{\rho^2}{\Delta_{r}}\ , \ 
g_{\theta\theta}=\frac{\rho^2}{\Delta_{\theta}}\ ,  
\end{eqnarray}
and the square of determinant of the metrics is  
$
\sqrt{-g}={\rho^2\sin{\theta}}/{\Xi}.
$

\subsection{Equations of scalar fields in Kerr-AdS spacetime} 

The action and the Lagrangian density of  
complex scalar field $\Phi$ with mass $\mu$ as a matter field 
in the Kerr-AdS spacetime is 
\begin{eqnarray}
I_{\rm M}&=& \int d^4 x 
\sqrt{-g} \, {L}_{\rm M} \ , \\ 
{L}_{\rm M}&=&
-g^{\mu\nu} \partial_{\mu}\Phi^* (x)\partial_{\nu}\Phi(x)
-{\mu}^2\Phi^*(x)\Phi(x) -\xi R\Phi^*\Phi \ ,
\end{eqnarray}
where the non-minimal coupling constant is denoted by $\xi$ and 
the scalar curvature takes the value $R=-12/\ell^2$.  
Field equations of scalar fields is 
\begin{eqnarray}
\left(\frac{1}{\sqrt{-g}}
\partial_{\mu}\sqrt{-g}g^{\mu\nu}\partial_{\nu}
 - \bar{\mu}^2 \right)\Phi = 0 \ , 
\end{eqnarray}
where $\bar\mu^2=\mu^2-12\xi/\ell^2$ denotes the effective mass  
of scalar field, which is assumed to take a non-negative value.
The contravariant metrics of the Kerr-AdS spacetime are       
\begin{eqnarray}
g^{tt}&=&
=\frac{1}{\rho^2}\left(\frac{a^2\sin^2{\theta}}{\Delta_{\theta}}
-\frac{(r^2+a^2)^2}{\Delta_{r}}\right) \ ,\  
g^{t\phi}
=\frac{a\Xi}{\rho^2}\left(\frac{1}{\Delta_{\theta}}
-\frac{r^2+a^2}{\Delta_{r}}\right) \ ,\nonumber\\
g^{\phi\phi}&=& 
=\frac{\Xi^2}{\rho^2}\left(\frac{1}{\Delta_{\theta}\sin^2{\theta}}
-\frac{a^2}{\Delta_{r}}\right) \ ,\ 
g^{rr}=\frac{1}{g_{rr}}\ , \ g^{\theta\theta}=\frac{1}{g_{\theta\theta}}\ .
\end{eqnarray}
One of zeros of $1/g^{tt}$ and $g^{rr}$ 
denote the horizons of black hole, namely,   
\begin{eqnarray}
\Delta_{r}={(r-r_{-})(r-r_{+})(r-r_{\ell})(r-r_{\ell}^{*})}/{\ell^2}\ , 
\label{horizon001}
\end{eqnarray}
where $r_{-}\, , r_{+}$ take real numbers,  
$r_{\ell}\, , r_{\ell}^{*}$ take complex numbers, which are assigned such that 
$r_{\pm}\rightarrow M\pm \sqrt{M^2-a^2}$,  
$r_{\ell}\rightarrow i\ell$ and $r_{\ell}^{*}\rightarrow -i\ell$
in the limit $\ell \rightarrow \infty \, (\Lambda \rightarrow 0)$.  
The event horizon corresponds to $r_{+}$. 

As the background Kerr-AdS metrics do not depend on time and azimuthal 
angle variables, 
the scalar field solution is put in the form of separation of 
variables: 
\begin{eqnarray}\label{separation001}
	\Phi 
=\frac{1}{\sqrt{2\pi}}{\rm e}^{-i\omega t}
{\rm e}^{i m \phi}S(\theta) R(r) \ ,   
\end{eqnarray}
where the frequency and azimuthal angular momentum are denoted by 
$\omega$ (complex value in general) and $m$ (integer value). 
Field equations for angular and radial parts become in the form:
\begin{eqnarray}
\left(
\frac{
\partial_{\theta}\sin{\theta}\Delta_{\theta}\partial_{\theta}}{\sin{\theta}}
-\frac{(a\omega\sin{\theta}-\Xi m/\sin{\theta})^2}{\Delta_{\theta}}
-\bar{\mu}^2a^2\cos^2{\theta}+\lambda \right)S(\theta)&=&0 
\label{separationangle001}\\
\left(
\partial_{r}\Delta_{r}\partial_{r} 
+\frac{((r^2+a^2)\omega-\Xi am)^2}{\Delta_{r}}
-\bar{\mu}^2r^2 -\lambda
\right)R(r)&=&0 , 
\label{separationradial001}
\end{eqnarray}
where $\lambda$ denotes the separation parameter  
\footnote{
It should be noted that 
field equations for angular and radial parts 
of Eqs.(7)-(8) in the paper by  Cardoso et al. \cite{cardoso001} 
(similarly in \cite{cardoso002})
are different from ours of Eqs.
(\ref{separationangle001})-(\ref{separationradial001}) 
though the same metric notation is used:   
Eqs.(2)-(3) in \cite{cardoso001} and 
Eqs.(\ref{line-element})-(\ref{metric}) in our paper. 
Their field equations cannot derive 
the general form of the Klein-Gordon inner product relations 
of Eq.(\ref{orthonormal:002}) nor 
the current conservation of particle number of Eq.(\ref{current-cons}).}.

\section{Normal modes of scalar fields}
\setcounter{equation}{0}
In this section, we consider the 
eigenstate problem of normal modes for scalar fields in Kerr-AdS
spacetime, 
which provides to analyze the super-radiant instability problem 
in the following sections. 
The boundary conditions on field equations are imposed  
and orthonormal relations among eigenfunctions for normal modes 
will be derived. 
The final part of this section, 
scalar fields are quantized and conserved quantities are obtained. 

\subsection{Boundary conditions } 
For the radial wave function, 
the Dirichlet boundary condition is imposed at infinity, 
because the spacetime is asymptotically AdS spacetime,   
and the Dirichlet or the Neumann boundary condition is imposed 
at the horizon to obtain normalized states: 
\begin{eqnarray}\label{normalboundary001}
&&R(r)\rightarrow 0 \ \ \  \mbox{\rm for} \ \ \ r\rightarrow \infty\ ,
\nonumber\\
&&R(r)=0 \ \ \mbox{\rm or}\ \ \frac{d}{dr}R(r)=0
\ \ \ \mbox{\rm for} \ \ \ r=r_{+}\ .
\end{eqnarray}
For the azimuthal angular function, 
the periodic boundary condition is imposed at $\phi=0$ and $2\pi$, 
which is satisfied for integer values of $m$.
For the polar angular function, 
the Dirichlet or the Neumann boundary condition is imposed 
at $\theta=0$ and $\pi$.

\subsection{Orthonormal relations }

From field equations 
of Eqs.(\ref{separationradial001}) and (\ref{separationradial001}), 
two identity equations in bi-linear forms of fields are obtained:
\begin{eqnarray}
\int^{\pi}_{0} d\theta \sin{\theta}\left(
(\omega^{*}-\omega{'})
\frac{(\omega^{*}+\omega{'})a^2\sin^2{\theta}-2ma\Xi}{\Delta_{\theta}}
-(\lambda^{*}-\lambda')\right)\nonumber\\
\times 
S^{*}_{\omega,m,\lambda}(\theta)S_{\omega',m,\lambda'}(\theta)=0\ , 
\end{eqnarray}
for angular part and 
\begin{eqnarray}
\int^{\infty}_{r_{+}} dr \left(
(\omega^{*}-\omega{'})
\frac{(\omega^{*}+\omega{'})(r^2+a^2)^2-2ma\Xi(r^2+a^2)}{\Delta_{r}}
-(\lambda^{*}-\lambda')\right)\nonumber\\
\times 
R^{*}_{\omega,m,\lambda}(r)R_{\omega',m,\lambda'}(r)=0\ ,
\end{eqnarray}
for radial part. From these identity equations, 
relations for the product of bi-linear forms are obtained:  
\begin{eqnarray}
(\omega^*-\omega')X_{(\omega,\omega',\lambda,\lambda')}=
(\lambda^*-\lambda')X_{(\omega,\omega',\lambda,\lambda')}=0 \ ,
\end{eqnarray}
where $X_{(\omega,\omega',\lambda,\lambda')}$ is defined as
\begin{eqnarray}
X_{(\omega,\omega',\lambda,\lambda')}&:=&
\int^{\pi}_{0} d\theta \int^{\infty}_{r_{+}} dr \frac{\sqrt{-g}}{\rho^2}
\nonumber\\
&&\times 
\left((\omega^{*}+\omega{'})\left( 
-\frac{a^2\sin^2{\theta}}{\Delta_{\theta}}+\frac{(r^2+a^2)^2
}{\Delta_{r}}\right)
+2ma\Xi\left(
\frac{1}{\Delta_{\theta}}-\frac{r^2+a^2}{\Delta_{r}}\right)
\right)
\nonumber\\
&&\times 
S^{*}_{\omega,m,\lambda}(\theta)S_{\omega',m,\lambda'}(\theta)
R^{*}_{\omega,m,\lambda}(r)R_{\omega',m,\lambda'}(r)\ . 
\end{eqnarray} 
According to the usual eigenvalue problems, 
the following cases are considered: 
\begin{itemize}
\item[(i)]
If $\omega=\omega'$ and $\lambda=\lambda'$, real values of frequencies and 
separation parameters are obtained: $\omega^*=\omega$ and
 $\lambda^*=\lambda$, 
because $X_{\omega,\omega,\lambda,\lambda}\neq 0$.
\item[(ii)]
If $\omega\neq\omega'$ and $\lambda\neq\lambda'$, 
orthogonal relations among different frequencies and 
separation parameters are obtained: 
$X_{\omega,\omega',\lambda,\lambda'}= 0$.
\end{itemize}
Combining (i) and (ii), orthonormal relations 
among angular and radial eigenfunctions for same azimuthal 
angular momenta $m$
are obtained:$\ 
X_{\omega,\omega',\lambda,\lambda'}
= \delta_{\omega,\omega'}\delta_{\lambda,\lambda'}, $
or in an explicit form  
\footnote{In deriving Eq.(\ref{ortho-ang-rad:001}), 
the following relation is used: 
\begin{eqnarray}
-g^{tt}(\omega^{*}+\omega{'}) 
+2g^{t\phi}m =
 \frac{(\omega^{*}+\omega{'})}{\rho^2}\left( 
-\frac{a^2\sin^2{\theta}}{\Delta_{\theta}}+\frac{(r^2+a^2)^2
}{\Delta_{r}}\right)
+\frac{2ma\Xi}{\rho^2}
\left(
\frac{1}{\Delta_{\theta}}-\frac{r^2+a^2}{\Delta_{r}}
\right). \nonumber
\end{eqnarray}
},   
\begin{eqnarray}
&&\int^{\pi}_{0} d\theta \int^{\infty}_{r_{+}} dr {\sqrt{-g}}
\left(-g^{tt}(\omega^{*}+\omega{'}) 
+2g^{t\phi}m \right)
\nonumber\\ && \times
S^{*}_{\omega,m,\lambda}(\theta)S_{\omega',m,\lambda'}(\theta)
R^{*}_{\omega,m,\lambda}(r)R_{\omega',m,\lambda'}(r)
=\delta_{\omega,\omega'}\delta_{\lambda,\lambda'}\ . 
\label{ortho-ang-rad:001}
\end{eqnarray} 
Similarly orthogonal relations  
for positive and negative  
values of azimuthal angular momenta $m,-m$ are obtained:  
\begin{eqnarray}
\label{ortho-ang-rad:002}
&&\int_{r_{+}}^{\infty} dr \int_{0}^{\pi} d\theta \sqrt{-g}
\left(-g^{tt}(\omega-\omega')+2g^{t\phi}m \right)\nonumber\\
&&
\times S_{\omega,m,\lambda}(\theta)S_{\omega',-m,\lambda'}(\theta)
R_{\omega,m,\lambda}(r)R_{\omega',-m,\lambda'}(r)=0 \ .  
\end{eqnarray}

The full eigenfunctions are defined:   
\begin{eqnarray}\label{generaleigenfunction001}
f_{\omega,m,\lambda}:=\frac{1}{\sqrt{2\pi}}
{\rm e}^{-i\omega t}{\rm e}^{im\phi}
S_{\omega,m,\lambda}(\theta)R_{\omega,m,\lambda}(r)\ , 
\end{eqnarray}
and the full orthonormal relations are obtained 
from Eqs.(\ref{ortho-ang-rad:001})-(\ref{ortho-ang-rad:002}) as 
\begin{eqnarray}
\int_{\Sigma} dr d\theta d\phi \sqrt{-g}\, 
(-g^{tt}(\omega+\omega')+g^{t\phi}(m+m'))\,
f_{\omega,m,\lambda}^{*}f_{\omega',m',\lambda'}\ 
&=&\delta_{\omega,\omega'}\delta_{m,m'}\delta_{\lambda,\lambda'}\ , \nonumber\\
 \int_{\Sigma} dr d\theta d\phi  \sqrt{-g}\, 
(-g^{tt}(\omega-\omega')+g^{t\phi}(m-m'))\,
f_{\omega,m,\lambda}f_{\omega',m',\lambda'}\ 
&=&0 \ , \label{orthonormal:001}
\end{eqnarray}
where the integration region $\Sigma$ is   
$0\leq\phi<2\pi$, $0\leq\theta<\pi$ and 
$r_{+}+\epsilon \leq r <\infty$.   
The cutoff parameter $\epsilon$ is introduced 
to regularize the divergent integration region 
due to the factor $g^{tt}$, which is  
considered in the brick wall model \cite{thooft001}. 
In the following, the parameter $\epsilon$ is omitted to write explicitly 
but it is understood to be recovered if necessary. 

The general Klein-Gordon inner product is introduced as a compact notation:  
\begin{eqnarray}
<A\ ,\ B>:=
\int_{\Sigma}d^3x 
\sqrt{-g}(-ig^{t\nu})
\left(A^{*}(t,x)\partial_{\nu}B(t,x)
-\partial_{\nu}A^{*}(t,x)B(t,x)\right)\ ,
\end{eqnarray}
where space coordinates are denoted as $x=(r,\theta,\phi)$ 
and time coordinate as $t$.  
The general form of orthonormal relations are written as   
\begin{eqnarray}
<f_{\alpha}\ ,\ f_{\alpha'}>&=&-<f_{\alpha}^{*}\ ,\ f_{\alpha'}^{*}>
=\delta_{\alpha,\alpha'}^{(3)}\nonumber \\
<f_{\alpha}^{*}\ ,\ f_{\alpha'}>&=&<f_{\alpha}\ ,\ f_{\alpha'}^{*}>
=0 \ , \label{orthonormal:002}
\end{eqnarray}
where the eigenvalues for normal modes are denoted:   
$\alpha:=(\omega,m,\lambda)$, 
$\alpha':=(\omega',m',\lambda')$ and $\delta_{\alpha,\alpha'}^{(3)}
:=\delta_{\omega,\omega'}\delta_{m,m'}\delta_{\lambda,\lambda'}$.
It is worthwhile to note that 
orthonormal relations hold in product forms of angular and 
radial parts (\ref{ortho-ang-rad:001})-(\ref{ortho-ang-rad:002})
but not in separate forms. It is also note that 
general orthonormal relations (\ref{orthonormal:002}) 
are the same as those of the general 
Klein-Gordon inner product relations \cite{birrell001}.

\subsection{Normal mode expansion and quantization}
The canonical momentum conjugate to scalar field is given by  
\begin{eqnarray}
\Pi:=\frac{\partial {L}_{\rm M}}{\partial \, \partial_{t}{\Phi}}
=-g^{t\nu}\partial_{\nu}\Phi^{\dagger}
=-(g^{tt}\partial_{t}{\Phi^{\dagger}}+g^{t\phi}\partial_{\phi}\Phi^{\dagger})
\ , 
\end{eqnarray}
where $\dagger$ denotes the Hermitian conjugate operation. 
Scalar fields and conjugate momenta are expressed in the normal 
mode expansion as  
\begin{eqnarray}
\Phi(t,x)&=&\sum_{\alpha}\left( 
a_{\alpha}f_{\alpha}(t,x)+b_{\alpha}^{\dagger}f_{\alpha}^{*}(t,x)
\right)
\ , \nonumber \\
\Pi(t,x)&=&-i
\sum_{\alpha}(g^{tt}\omega-g^{t\phi}m)
(a_{\alpha}^{\dagger}f_{\alpha}^{*}(t,x)-b_{\alpha}f_{\alpha}(t,x))
\ . 
\label{nm-expansion001}
\end{eqnarray}
The equal time commutation relations  
among fields and their momenta are imposed:   
\begin{eqnarray}
\left[\Phi(t,x),\Pi(t,x')\right]
=\left[\Phi^{\dagger}(t,x),
\Pi^{\dagger}(t,x')\right]=\frac{i}{\sqrt{-g}}\delta^{(3)}(x-x')\ , 
\end{eqnarray}
and others are zeros. From these commutation relations, 
commutation relations among annihilation and creation operators 
are derived: 
\begin{eqnarray}
[a_{\alpha},a_{\alpha'}^{\dagger}]
=[b_{\alpha},b_{\alpha'}^{\dagger}]
=\delta^{(3)}_{\alpha ,\alpha'}\ , 
\end{eqnarray} 
and others are zeros,  
where the completeness relations are used:
\begin{eqnarray}
\sum_{\alpha}(-g^{tt}\omega+g^{t\phi} m)
\left(f_{\alpha}(t,x)f_{\alpha}^{*}(t,x')
+f_{\alpha}^{*}(t,x)f_{\alpha}(t,x')\right)
&=&\frac{1}{\sqrt{-g}}\delta^{(3)}(x-x')\ , \nonumber \\
\sum_{\alpha}
\left(f_{\alpha}(t,x)f_{\alpha}^{*}(t,x')
-f_{\alpha}^{*}(t,x)f_{\alpha}(t,x')\right)&=&0\ .
\end{eqnarray}
It is noted that the commutation relations 
in the Kerr-AdS spacetime are the same form as those 
in the Minkowski spacetime except for the metric determinant factor 
$1/\sqrt{-g}$.  
  
\subsection{Energy and angular momentum}

Because the metrics are independent of time and azimuthal angle, 
two Killing vectors exist: $\ 
\xi_{(t)}^{\mu}={\partial}/{\partial t}
\ , \ \xi_{(\phi)}^{\mu}={\partial}/{\partial \phi}.$
Defining the energy-momentum tensor  
\begin{eqnarray}
T_{\mu\nu}&:=&-\frac{2}{\sqrt{-g}}\frac{\delta I_{\rm M}}{\delta g^{\mu\nu}}
\nonumber\\
&=&\partial_{\mu}\Phi^{\dagger}\partial_{\nu}\Phi
+\partial_{\nu}\Phi^{\dagger}\partial_{\mu}\Phi
-g_{\mu\nu}(g^{\alpha\beta}\partial_{\alpha}\Phi^{\dagger}\partial_{\beta}\Phi
+\bar{\mu}^2\Phi^{\dagger}\Phi)\ ,
\end{eqnarray}
local conservation laws hold for each Killing vector 
\begin{eqnarray}
\partial_{\nu}(\sqrt{-g}\,\xi_{(i)}^{\mu} T_{\mu}^{\nu})=0 \ \ , \ \ 
{\mbox{ for}}\ \ i=t, \phi \ .
\end{eqnarray}
Corresponding conservative quantities are 
energy and angular momentum:   
\begin{eqnarray}
E&=&- \int_{\Sigma} 
d^3x \sqrt{-g}\,\xi_{(t)}^{\mu}T_{\mu}^{t}\nonumber\\
&=& \int_{\Sigma} d^3x\sqrt{-g}(
-g^{tt}\partial_{t}\Phi^{\dagger}\partial_{t}\Phi
+g^{\phi\phi}\partial_{\phi}\Phi^{\dagger}\partial_{\phi}\Phi
\nonumber\\
&&\hspace{5em}+g^{rr}\partial_{r}\Phi^{\dagger}\partial_{r}\Phi 
+g^{\theta\theta}\partial_{\theta}\Phi^{\dagger}\partial_{\theta}\Phi)
\\
L&=& \int_{\Sigma} 
d^3x\sqrt{-g}\,\xi_{(\phi)}^{\mu}T_{\mu}^{t}\nonumber\\
&=& \int_{\Sigma} d^3x
\sqrt{-g}(g^{tt}(\partial_{t}\Phi^{\dagger}\partial_{\phi}
\Phi
+\partial_{\phi}\Phi^{\dagger}\partial_{t}\Phi)
+2g^{t\phi}\partial_{\phi}\Phi^{\dagger}\partial_{\phi}\Phi) \ . 
\end{eqnarray}
They are expressed by creation and annihilation 
operators using the normal node expansion as 
\begin{eqnarray}
E=\sum_{\alpha}\omega(a_{\alpha}^{\dagger}a_{\alpha}
+b_{\alpha}b_{\alpha}^{\dagger})\ , \
L=\sum_{\alpha}m(a_{\alpha}^{\dagger}a_{\alpha}
+b_{\alpha}b_{\alpha}^{\dagger})\ . 
\end{eqnarray}
In combining these conserved quantities, 
the effective energy is defined, 
which is the energy taking into the rotation effect on the horizon, as 
\begin{eqnarray}
E-\Omega_{\rm H}L
&=& \sum_{\alpha}(\omega-\Omega_{\rm H}m)(a_{\alpha}^{\dagger}a_{\alpha}
+b_{\alpha}b_{\alpha}^{\dagger})\ , 
\label{eqn:5012}
\end{eqnarray}
where the angular velocity on the horizon is defined: 
\begin{eqnarray}
\label{velocity001}
\Omega_{\rm H}:=\left.\frac{g^{tt}}{g^{t\phi}}\right|_{r=r_{+}}
=\frac{a\Xi}{r^2_{+}+a^2} \ .
\end{eqnarray}
From the expression in Eq.(\ref{eqn:5012}), 
the effective energy $E-\Omega_{\rm H }L$ 
is positive definite if $0<\omega-\Omega_{\rm H}m$. 
This condition is important on the super-radiant instability and 
the definition of the statistical mechanics 
for scalar fields in Kerr-AdS spacetime, which will be studied  
in section 5 and in appendix. 
 
\section{Quasi-normal modes of scalar fields}
\setcounter{equation}{0}
In this section, we consider quasi-normal modes 
of scalar fields in Kerr-AdS spacetime to extend  
normal modes results in the previous section. 
Quasi-normal modes are also important in analyzing the stability 
problem. 
Our treatment of boundary conditions and initial conditions 
for quasi-normal modes is natural extension of 
the case for normal modes. 

\subsection{Boundary conditions and initial condition}
We impose the Dirichlet boundary condition at infinity and  
the ingoing (into black holes) 
boundary condition at horizon 
on the radial wave function for quasi-normal modes: 
\begin{eqnarray}
\label{quasinormalboundary001}
&&R(r)\rightarrow \left\{ \begin{array}{ll}
0 &  \mbox{\rm for} \ \ \ r\rightarrow \infty \\
\exp{(-i(\omega-\Omega_{\rm H}m)r_{*})} & \mbox{\rm for} \ \  \ r=r_{\rm +}\ ,
\end{array}\right.
\end{eqnarray}
where $r_{*}$ denotes the Regge-Wheeler tortoise coordinate: 
\begin{eqnarray}
\label{tortoise001}
r_{*}:=
\int_{\infty}^{r} dr \frac{r^2+a^2}{\Delta_{r}}
\simeq 
\left\{ \begin{array}{ll}
\displaystyle
\frac{\ell^2(r_{+}^2+a^2)\ln(r-r_{+}) }
{(r_{+}-r_{-})(r_{+}-r_{\ell})(r_{+}-r_{\ell}^{*})}
 & \mbox{ for $r \simeq r_{+}$} \\
\displaystyle
-\frac{\ell^2}{r} & \mbox{for $r \rightarrow \infty$}
\end{array}
\right. \ . 
\label{RW-coordinate}
\end{eqnarray}
As the initial condition for orthonormal relations of quasi-normal
modes, we impose them to be coincident with those for normal modes in 
Eq.(\ref{orthonormal:001}) or Eq.(\ref{orthonormal:002})  
in case of real values of $\omega$ and $\lambda$.

\subsection{Quasi-orthonormal relations}
In order to obtain quasi-orthonormal relations for quasi-normal modes, 
we follow calculations of orthonormal relations for normal modes 
in keeping the boundary term in Eq.(\ref{ortho-ang-rad:001}). 
The bi-linear identity is obtained with the boundary contribution as 
\begin{eqnarray}
(\omega^{*}-\omega')\int_{r_{+}}^{\infty}dr\int_{0}^{\pi} d\theta \sqrt{-g}
\left(
-g^{tt}(\omega^{*}+\omega')+2g^{t\phi}m
\right)S_{\bar{\alpha}}^{*}(\theta)R_{\bar{\alpha}}^{*}(r)
S_{\bar{\alpha}'}(\theta)R_{\bar{\alpha}'}(r)\nonumber \\
=\int_{0}^{\pi} d\theta \frac{\sin{\theta}}{\Xi}
S_{\bar{\alpha}}^{*}(\theta)S_{\bar{\alpha}'}(\theta)
\Delta_{r}\left. \left(
R_{\bar{\alpha}}^{*}(r)\frac{d}{dr}R_{\bar{\alpha}'}(r)
-\frac{d}{dr}R_{\bar{\alpha}}^{*}(r)R_{\bar{\alpha}'}(r)
\right)\right|_{r_{+}}^{\infty}\ ,
\label{qnm-bilinear}
\end{eqnarray}
where $\bar{\alpha}=(\omega,m,\lambda)$ and 
$\bar{\alpha}'=(\omega',m,\lambda')$ (same $m$)
with complex values of $\omega$ and $\lambda$. 
Using the general form of eigenfunctions $f_{\alpha}$
in Eq.(\ref{generaleigenfunction001}),  
the bi-linear identity in Eq.(\ref{qnm-bilinear}) is rewritten 
in the form  
\begin{eqnarray}
&&\frac{d}{dt}\int_{\Sigma} dr d\theta d\phi \sqrt{-g}g^{t\nu}
(f_{\alpha}^{*}\partial_{\nu}f_{\alpha'}
-\partial_{\nu}f_{\alpha}^{*}f_{\alpha'}
) \nonumber\\
&=&
\left.  \int_{0}^{\pi}d\theta\int_{0}^{2\pi}
d\phi \sqrt{-g}g^{rr}
(f_{\alpha}^{*}\partial_{r}f_{\alpha'}
-\partial_{r}f_{\alpha}^{*}f_{\alpha'}
)\right|_{r_{+}}^{\infty}\ ,
\label{qnm-bilinear-general}
\end{eqnarray} 
with $\alpha=(\omega,m,\lambda)$ and $\alpha'=(\omega',m',\lambda')$.  
Integrating this equation, the general form of the identity is obtained:
\begin{eqnarray}
&-i&\int_{\Sigma} dr d\theta d\phi \sqrt{-g}g^{t\nu}
(f_{\alpha}^{*}\partial_{\nu}f_{\alpha'}
-\partial_{\nu}f_{\alpha}^{*}f_{\alpha'}
) \nonumber\\
&=&
i\left. \int_{0}^{t}dt \int_{0}^{\pi}d\theta\int_{0}^{2\pi}
d\phi \sqrt{-g}g^{rr}
(f_{\alpha}^{*}\partial_{r}f_{\alpha'}
-\partial_{r}f_{\alpha}^{*}f_{\alpha'}
)\right|_{r_{+}}^{\infty}
+{ C_{\alpha,\alpha'}}\ ,
\label{qnm-int-identity}
\end{eqnarray}
where the integration constant is 
determined by the initial value of bilinear forms:  
\begin{eqnarray}
C_{\alpha,\alpha'}:= 
\left. 
-i\int_{\Sigma} dr d\theta d\phi \sqrt{-g}g^{t\nu}
(f_{\alpha}^{*}\partial_{\nu}f_{\alpha'}
-\partial_{\nu}f_{\alpha}^{*}f_{\alpha'})
\right|_{t=o}\ ,
\label{initial001}
\end{eqnarray}
with $C_{\alpha,\alpha}=1$ as the normalization condition for
quasi-normal modes. 
Defining the quasi-inner product by 
\begin{eqnarray}\label{quasi-orthonormal:001}
&&<<A,B>>:=
\int_{\Sigma} d^3x \sqrt{-g}(-ig^{t\nu})(A^{*}(t,x)\partial_{\nu}B(t,x)
-\partial_{\nu}A^{*}(t,x)B(t,x))\nonumber\\
&&+ \left. 
\int_{0}^{t}dt \int_{0}^{\pi}d\theta\int_{0}^{2\pi} d\phi \sqrt{-g}(-ig^{rr})
(A^{*}(t,x)\partial_{r}B(t,x)-\partial_{r}A^{*}(t,x)B(t,x) )
\right|_{r_{+}}^{\infty}\ , \nonumber\\
\ \ \ 
\end{eqnarray}
the quasi-orthonormal relations are expressed: 
\begin{eqnarray}\label{quasi-orthonormal:002}
<<f_{\alpha},f_{\alpha'}>>=C_{\alpha,\alpha'}\ .
\label{qnm-ortho-c}
\end{eqnarray} 
Similar relations are also obtained:
\begin{eqnarray}
<<f_{\alpha}^{*},f_{\alpha'}^{*}>>=C_{\alpha,\alpha'}^{*}
\ , \ 
<<f_{\alpha}^{*},f_{\alpha'}>>=D_{\alpha,\alpha'}\ , \  
<<f_{\alpha},f_{\alpha'}^{*}>>=D_{\alpha,\alpha'}^{*}\ ,
\label{qnm-ortho-d}
\end{eqnarray}   
where the other integration constants are defined:  
\begin{eqnarray}
D_{\alpha,\alpha'}:= 
\left. 
-i\int_{\Sigma} dr d\theta d\phi \sqrt{-g}g^{t\nu}
(f_{\alpha}\partial_{\nu}f_{\alpha'}
-\partial_{\nu}f_{\alpha}f_{\alpha'})
\right|_{t=o}\ .
\end{eqnarray}

Here we consider a single quasi-normal mode 
of $0<{\rm  Re}\,\omega-\Omega_{\rm H}m$, which is 
normalized to one at the initial time  
\begin{eqnarray}
\int d^3 x \sqrt{-g}(-2)(g^{tt}{\rm
 Re}\,\omega-g^{t\phi}m)|f_{\alpha}|^2=1  \ \ \ \mbox{at} \ \ \ t=0 
\ ,
\end{eqnarray}
according to the initial condition in Eq.(\ref{initial001}). 
This mode decreases with time to tend to zero at $t=\infty$. 
On the other hand, the boundary term increases from zero to one 
as the flux flows into black holes, which gives the sum rule for the
imaginary part of the frequency: 
\begin{eqnarray}
\label{sumrule001}
{\rm Im}\, \omega&=&\left. -\frac{1}{2}\int_{0}^{\pi}d\theta \int_{0}^{2\pi}
 d\phi \sqrt{-g}(-ig^{rr})
(f_{\alpha}^{*}\partial_{r}f_{\alpha}
-\partial_{r}f_{\alpha}^{*}f_{\alpha})\right|_{r_{+}}^{\infty} 
\nonumber\\
&=&-\int_{0}^{\pi} d\theta \frac{\sin{\theta}}{\Xi}(r_{+}^2+a^2)
({\rm Re}\,\omega-\Omega_{\rm H}m)|S_{\alpha}(\theta)R_{\alpha}(r_{+})|^2
\ . 
\end{eqnarray}
This sum rule means that the total flux flowing into black holes 
determines the imaginary part of the quasi-normal mode. 
Wee see that ${\rm Im}\, \omega$ is negative for 
$0<{\rm Re}\, \omega-\Omega_{\rm H}m$. 
We will discuss this point again in the connection with the 
non-existence of zero mode in section 5. 
It is worthwhile to note that 
the conditions $0<{\rm  Re}\,\omega-\Omega_{\rm H}m$ and $0<{\rm
Im}\,\omega$ are supported by the co-rotating frame consideration: 
Eqs.(\ref{real-part})-(\ref{imaginary-part}) in appendix. 

These quasi-inner product relations for quasi-normal modes 
are the extension of the inner products for normal modes 
to include boundary terms. 
It is noted that the quasi-orthonormal relations of 
Eqs.(\ref{qnm-ortho-c})-(\ref{qnm-ortho-d}) are consistent with 
the current conservation of particle number:
\begin{eqnarray}
\partial_{\mu}
\left(-ig^{\mu\nu}(\Phi^{*}\partial_{\nu}\Phi-\partial_{\nu}\Phi^{*}
\Phi)\right)=0\ .\label{current-cons}
\end{eqnarray}
Related to quasi-orthonormal relations, 
it is also noted that the completeness of quasi-normal modes 
in the sense of normal modes was studied by 
Kokkotas and Scmidt \cite{kokkotas001} and by 
Price and Husain \cite{price001}.

\subsection{Flux of energy and angular momentum}

The energy and angular momentum of scalar fields are conserved in time 
for normal modes but 
they vary with time for quasi-normal modes. 
The time derivative of energy and momentum are expressed by the flux of them: 
\begin{eqnarray}
\frac{dE}{dt}&=&
\left. \int_{0}^{\pi} d\theta \int_{0}^{2\pi}d\phi 
\sqrt{-g}\xi_{(t)}^{\mu}T_{\mu}^{r}
\right|_{r_{+}}^{\infty} \nonumber\\
&=&\left. \int_{0}^{\pi} d\theta \int_{0}^{2\pi}d\phi \sqrt{-g}g^{rr}
(\partial_{t}\Phi^{\dagger}\partial_{r}\Phi+
\partial_{r}\Phi^{\dagger}\partial_{t}\Phi)
\right|_{r_{+}}^{\infty} , \\
\frac{dL}{dt}&=&
-\left. \int_{0}^{\pi} d\theta \int_{0}^{2\pi}d\phi \sqrt{-g}\xi_{(\phi)}^{\mu}T_{\mu}^{r}
\right|_{r_{+}}^{\infty}\nonumber\\
&=&-\left. \int_{0}^{\pi} d\theta \int_{0}^{2\pi}d\phi \sqrt{-g}g^{rr}
(\partial_{\phi}\Phi^{\dagger}\partial_{r}\Phi+
\partial_{r}\Phi^{\dagger}\partial_{\phi}\Phi)
\right|_{r_{+}}^{\infty}\ .
\end{eqnarray}
From these expressions, 
the time derivative of the effective energy  
is shown to be negative definite as 
\begin{eqnarray}\label{flux001}
\frac{d(E-\Omega_{\rm H}L)}{dt}
&=&
-2\int_{0}^{\pi} d\theta \int_{0}^{2\pi}d\phi
\frac{\sin{\theta}(r_{+}^2+a^2)}{\Xi}
\left|
(\partial_{t}+\Omega_{\rm H}\partial_{\phi})\Phi
\right|^{2}
\ .\label{energy-flux}
\end{eqnarray}
 
For a single quasi-normal mode with 
$\alpha=(\omega,m,\lambda)$, the time derivative of the effective 
energy satisfies the relation 
\begin{eqnarray}
\frac{d(E_{\alpha}-\Omega_{\rm H}L_{\alpha})}{dt}=2{\rm Im}\, \omega
(E_{\alpha}-\Omega_{\rm H}L_{\alpha})\ ,
\label{flux002}
\end{eqnarray}
where 
\begin{eqnarray}
E_{\alpha}-\Omega_{\rm H}L_{\alpha}=
({\rm Re}\, \omega-\Omega_{\rm H}m)
\int_{\Sigma} d^3x \sqrt{-g}(-2)(g^{tt}{\rm Re}\,
\omega-g^{t\phi}m)|f_{\alpha}|^2
\ , \label{flux003}
\end{eqnarray}
which is positive definite for 
$0<{\rm Re}\, \omega-\Omega_{\rm H}m$.   
Together with the negative definiteness of 
$d(E-\Omega_{\rm H} L)/dt$ of Eq.(\ref{flux001}) and 
the positive definiteness of the effective energy 
of Eq.(\ref{flux003}) for 
$0<{\rm Re}\, \omega-\Omega_{\rm H}m$,  
${\rm Im}\, \omega$ is negative definite consistent with 
the sum rule of Eq.(\ref{sumrule001}).

\section{Zero modes and super-radiant modes}
\setcounter{equation}{0}
In this section, we study 
zero modes defined as
${\rm Re}\, \omega-\Omega_{\rm H}m=0$ 
$(-\infty <m< \infty)$ 
and super-radiant unstable modes  
 ${\rm Re}\, \omega-\Omega_{\rm H}m \leq 0$ 
$(0<{\rm Re}\,\omega$ and $ -\infty <m< \infty )$ 
relating to  
normal modes and quasi-normal modes studied in previous sections. 
Each mode is specified by a set of values; 
$\omega$, $m$ and $\lambda$, where $\omega$ and $\lambda$ 
can be complex values and $m$ takes a integer number. 
The separation parameter $\lambda$ is considered to be fixed value 
throughout in this section.
Some statements are provided in the following.
\begin{item}
\item[\bf Statement 1]
Boundary terms for zero modes vanish,  
and the frequency and the separation parameter of zero modes 
take real values.
\item[[Proof]]
The identity for the boundary term  
\begin{eqnarray}
&&-ig^{rr}\left(
f_{\alpha}^{*}(x)\partial_{r}f^{\alpha}(x)
-\partial_{r}f_{\alpha}^{*}(x) f^{\alpha}(x)
\right)\nonumber\\
&&=-2({\rm Re} \,\omega-\Omega_{\rm H}m)
\frac{r_{+}^2+a^2}{r_{+^2}+a^2\cos^2{\theta}}
f_{\alpha}^{*}(x) f^{\alpha}(x)\ \ {\rm at}\ \ r=r_{+}\ , 
\end{eqnarray}
shows that boundary contributions  
in Eqs.(\ref{qnm-bilinear}) 
and (\ref{qnm-bilinear-general}) 
vanish for zero modes 
regardless of boundary conditions.
The zero modes are special modes in the sense that  
they are not ingoing or outgoing modes (or oscillating modes)   
but the linear function with respect to the tortoise coordinate:
\begin{eqnarray}
R_{\rm zero}\simeq d_{1}+d_{2}r_{*}\ \ \ \mbox{\rm near} \ \ \ 
r\simeq r_{+} \ , 
\label{zm-radial}
\end{eqnarray}
where $d_{1}, d_{2}$ are integration constants. 
The frequency and the separation parameter for zero modes 
take real values 
according to the similar argument for normal modes in section 3. 

\item[\bf Statement 2] 
A mode of $(\omega,m)$ 
has its reflection symmetric partner mode $(-\omega,-m)$. 
The zero mode line $0={\rm Re}\,\omega-\Omega_{\rm H}m$ 
in $\omega-m$ plane is invariant under the reflection transformation. 
\item[[Proof]]
Field equations of Eqs.(\ref{separationangle001})
-(\ref{separationradial001}) 
are invariant under the reflection transformation:
$\omega\rightarrow -\omega, m \rightarrow -m$ 
and any solutions form pair modes: $(\omega,m)$ and
      $(-\omega,-m)$. 
The zero mode line $0={\rm Re}\,\omega-\Omega_{\rm H}m$ 
in $\omega-m$ plane is transformed 
into itself by the reflection transformation 
and is reflection symmetric.
\item[\bf Statement 3]
If a mode $(\omega,m)$ is a physical mode, 
its reflection partner $(-\omega,-m)$ is an unphysical mode. 
Zero modes do not exist as physical modes or unphysical modes. 
\item[[Proof]]
Generally if a mode $(\omega,m)$ is a physical mode described by 
a particle annihilation operator $a_{\alpha}$, 
its reflection partner $(-\omega,-m)$ is an unphysical mode 
described by an antiparticle creation operator $b_{\alpha}^{+}$ 
(see Eq.(\ref{nm-expansion001}))
according to the usual relativistic theory \cite{weinberg001}. 
The zero mode line $0={\rm Re}\,\omega-\Omega_{\rm H}m$ 
in $\omega-m$ plane cannot be divided into two parts, 
physical and unphysical modes, because the zero mode line  
is invariant under the reflection transformation.  

As another evidence for the non-existence of zero modes, 
the constant zero mode of the first term in 
Eq.(\ref{zm-radial}) (a candidate of zero modes) 
cannot be normalized because the  
suppression factor $r^2-r_{+}^2$ appears in the integrand of the 
normalization equation (\ref{ortho-ang-rad:001}): 
\begin{eqnarray}
\label{zeronorm001}
\int_{r_{+}}^{\infty} dr 
\int_{0}^{\pi}d\theta \sqrt{-g}\frac{-2g^{tt}\Xi a m (r^2-r_{+}^2)}
{(r_{+}^2+a^2)(r^2+a^2)}|S_{\rm zero}(\theta)R_{\rm zero}(r)|^2\simeq 1 
\ ,\
\end{eqnarray}
where the near horizon approximation is applied because of 
the enhanced factor $g^{tt}$ in the integrand. 
The second term in Eq.(\ref{zm-radial}) 
(another candidate of zero modes) do not satisfy the boundary
 conditions. 
Therefore any zero mode solutions 
cannot satisfy the boundary conditions or the normalization condition.

\item[\bf Statement 4]
The allowed Physical mode region is  
$0<{\rm Re}\,\omega-\Omega_{\rm H}m$ $({\rm Re}\,\omega<0$ 
or $0<{\rm Re}\,\omega$  and $-\infty<m<\infty)$.
\item[[Proof]]
Consider first the non-rotating case 
with zero black hole rotation parameter: $J=aM=0$. 
Physical modes are $0<{\rm Re}\,\omega$ $( -\infty < m <\infty)$ 
and unphysical modes are reflection symmetric modes:  
${\rm Re}\,\omega <0$ $(-\infty <m< \infty)$. 
We assume that physical modes are  
analytic with respect to the rotation parameter $J$.  
Consider next the rotating case. 
Physical modes shift from $0< {\rm Re}\,\omega$ to 
$0< {\rm Re}\,\omega-\Omega_{\rm H}m$ $(-\infty <m<\infty)$ 
because any physical modes cannot cross the zero mode line during 
the change of the rotation parameter from $0=J$ to $J \neq 0$. 
It is noted that 
the value of $J$ is defined for the real value of horizon: $0<r_{-}<r_{+}$.  
We know that 
one of important roles of the zero mode line is to separate physical modes 
from unphysical modes. 
This statement is consistent with the 
co-rotating frame consideration of Eq.(\ref{real-part}) in appendix 
\footnote{
Related to the non-normalizability of zero modes (Statement 3) 
and the allowed physical mode region (Statement 4), 
orthonormal relation of Eq.(\ref{orthonormal:001}) 
or Eq.(\ref{orthonormal:002}) and quasi-orthonormal relations 
of Eqs.(\ref{qnm-ortho-c})-(\ref{qnm-ortho-d}) are recognized 
to be valid for physical modes: $0<{\rm Re}\,\omega-\Omega_{\rm H}m$.   
}.  
\item[\bf Statement 5]
The imaginary part of the frequency of quasi-normal modes 
become negative and scalar fields 
in Kerr-AdS spacetime is stable.
\item[[Proof]]
The sum rule for the imaginary part in Eq.(\ref{sumrule001}) and 
the effective energy flow in Eqs.(\ref{flux002})-(\ref{flux003}) 
combining the negative definite expression 
of the effective energy in Eq.(\ref{flux001}) 
show that the imaginary part of the frequency for zero modes is negative.
This statement is also consistent with the co-rotating frame 
consideration of Eq.(\ref{imaginary-part}) in the appendix.

\item[\bf Statement 6] 
The super-radiant instability 
for ${\rm Re}\,\omega-\Omega_{\rm H}m<0$ with $0<{\rm Re}\,\omega$ 
does not occur 
though  the stable 
super-radiant modes for $0<{\rm Re}\,\omega-\Omega_{\rm H}m$ 
with ${\rm Re}\,\omega<0$ exist as physical modes.
\item[[Proof]]
According to Statement 4, 
unstable super-radiant modes 
for ${\rm Re}\,\omega-\Omega_{\rm H}m<0$ with $0<{\rm Re}\,\omega$ 
are unphysical modes and their reflection partners 
$0<{\rm Re}\,\omega-\Omega_{\rm H}m$ with ${\rm Re}\,\omega<0$ 
are physical modes, which are considered as stable super-radiant modes. 
\item[\bf Statement 7] 
The statistical mechanics for scalar fields around Kerr-AdS spacetime 
is well-defined, namely, the partition function 
by Hartle and Hawking \cite{hawking002}
\begin{eqnarray} 
Z={\rm Tr} \exp{(-\beta_{\rm H}(E-\Omega_{\rm H}L))}\ ,
\end{eqnarray}
becomes well-defined. 
\item[[Proof]]
The effective energy in Eq.(\ref{eqn:5012}) 
becomes positive definite $0<E-\Omega_{\rm H}L$ 
for the allowed physical mode region    
$0<{\rm Re}\,\omega-\Omega_{\rm H}m$.  
This establishes the brick wall model 
for rotating black holes \cite{kenmoku001,kenmoku002}.
\end{item}

\section{Summary}
\setcounter{equation}{0}
We have studied normal modes, quasi-normal modes,  
zero modes and super-radiant modes for  
scalar fields in (3+1)-dimensional 
Kerr-AdS in order to make clear 
the scalar perturbation around rotating black hole spacetime. 

For normal modes, 
radial eigenfunctions and polar angle eigenfunctions  
are shown to satisfy the orthonormal relations in the product forms 
Eqs.(\ref{ortho-ang-rad:001})-(\ref{ortho-ang-rad:002}) 
but not in the separate forms.
Each normal mode is specified by a set of real values 
$(\omega,m,\lambda)$.  
Quantum scalar fields are expressed in the normal mode expansion 
with the creation and annihilation operators.
The effective energy $E-\Omega_{\rm H}L$ is shown to be positive definite 
for $0<\omega-\Omega_{\rm H}m$.

For quasi-normal modes,  
the quasi-orthonormal relations are obtained taking account of 
the boundary effect and 
are consistent with the current conservation of particle number. 
The sum rule for the imaginary part of frequency is obtained 
in Eq.(\ref{sumrule001}), which shows the negative value of 
${\rm Im}\,\omega$. 
The flux of the effective energy is shown to be negative definite 
in Eq.(\ref{flux001}). 
The imaginary part of the frequency is again to be negative 
using the relation of Eq.(\ref{flux002}) and 
the positive value of $E-\Omega_{\rm H}L$ 
in Eq.(\ref{flux003}) for $0<{\rm Re}\,\omega-\Omega_{\rm H}m$.  
The negative value of ${\rm Im}\,\omega$ indicates that 
the scalar fields in the Kerr-AdS spacetime is stable.  

For zero modes: $0=\omega-\Omega_{\rm H}m$, 
they are shown not to exist because 
the zero mode line in $0=\omega-m$ plane 
is invariant under the reflection transformation 
and cannot be identified as physical modes or unphysical modes.  
Other evidence for the non-existence of zero mode is that they  
cannot satisfy the boundary condition or the normalization condition.  
Physical modes are shown to be $0<\omega-\Omega_{\rm H}m$, 
because of the non-existence of zero modes and 
the analyticity of rotation parameter $J$. 
(The value of $J$ is allowed for 
well-defined values of the horizon $0<r_{-}<r_{+}$.) 
This fact implies that 
the important role of the zero mode line is to separate physical modes from 
unphysical modes. 

For super-radiant modes, 
unstable super-radiant modes ${\rm Re}\,\omega-\Omega_{\rm H}m<0$ with 
$0<{\rm Re}\,\omega$ do not occur but 
stable super-radiant modes 
$0<{\rm Re}\,\omega-\Omega_{\rm H}m$ with ${\rm Re}\,\omega<0$ can
occur. 

The results are consistent with orthonormal relations 
(Eq.(\ref{orthonormal:001})) or Eq.(\ref{orthonormal:002}), 
quasi-orthonormal relations Eqs.(\ref{qnm-ortho-c})-(\ref{qnm-ortho-d})  
and the co-rotating frame consideration:  
$0<{\rm Re}\,\omega-\Omega_{\rm H}m$ (Eq.(\ref{real-part})) and  
${\rm Im}\omega<0$ (Eq.(\ref{imaginary-part})) in appendix.  
The results for (3+1)-dimensional Kerr-AdS spacetime 
are also consistent with 
those for (2+1)-dimensional BTZ black hole spacetime 
\cite{kenmoku003,kenmoku001}. 
As our method does not essentially depend on spacetime dimensionality, 
the application to higher dimensional rotating black holes 
will be possible and interesting \cite{kenmoku005}.  
Explicit construction and numerical analysis of eigenvalue solutions 
for scalar fields in Kerr-AdS spacetime will be published in a separate
paper \cite{kenmoku004}. \\

\noindent
{\Large{\bf Acknowledgements}}\\

The author thanks Dr. Kazuyasu Shigemoto for useful discussions 
and advices. The author also thanks Prof. Haruhiko Terao 
for enough support to this work.


\appendix
\section{Co-rotating frame}
In this appendix,  
we consider the co-rotating coordinate, which 
diagonalize the metric in $t-\phi$ spacetime as  
\begin{eqnarray}
g_{tt}dt^2+g_{t,\phi}dtd\phi+g_{\phi,\phi}d\phi^2
=
g_{\hat{t} \tilde{t}}d\tilde{t}^2+g_{\tilde{\phi} \tilde{\phi}}d\tilde{\phi}^2
\ ,   
\end{eqnarray}
where the transformation is defined:  
\begin{eqnarray}
(d\tilde{t}, d\tilde{\phi})=(dt, d\phi)S\ ,
\end{eqnarray}
with the transformation matrix 
\begin{eqnarray}
S:=\frac{\sqrt{r^2+a^2}}{\rho}
\left(
\begin{array}{cc}
1 & -a\Xi/(r^2+a^2) \\
-a\sin^2{\theta}/\Xi & 1
\end{array}
\right)\ , 
\label{trans-matrix}
\end{eqnarray}
and the diagonal metrics are obtained:     
\begin{eqnarray}
\left(
\begin{array}{cc}
g_{\tilde{t}\tilde{t}} & 0 \\
0 & g_{\tilde{\phi}\tilde{\phi}}
\end{array}
\right)
=
\left(
\begin{array}{cc}
-\Delta_{r}/(r^2+a^2) & 0\\
0 & \Delta_{\theta}(r^2+a^2)\sin^2{\theta}/\Xi^2
\label{diag-diff-op}
\end{array}
\right)\, .
\end{eqnarray}
Under the transformation of Eq.(\ref{trans-matrix}), 
the Klein-Gordon operator for $t-\phi$ components 
is also diagonalized: 
\begin{eqnarray}
g^{tt}\partial_{t}^2+2t^{t\phi}\partial_{t}\partial_{\phi}
+g^{\phi\phi}\partial_{\phi}^2=
g^{\tilde{t}\tilde{t}}\partial_{\tilde{t}}^2
+g^{\tilde{\phi}\tilde{\phi}}\partial_{\tilde{\phi}}^2 \ ,
\end{eqnarray}
with the diagonal inverse metrics:
$g^{\tilde{t}\tilde{t}}=1/g_{\tilde{t}\tilde{t}}\ , \ 
g^{\tilde{\phi}\tilde{\phi}}=1/g_{\tilde{\phi}\tilde{\phi}}$.
The diagonal form of the Klein-Gordon operator  
leads to the field equations 
in separation of variables 
of Eqs.(\ref{separationangle001})-(\ref{separationradial001}).

A invariant is formed by the combination of coordinates and
differential operators: 
\begin{eqnarray}
dt\,\partial_{t}+d\phi\, \partial_{\phi}
=d\tilde{t}\, \partial_{\tilde{t}}+\tilde{\phi}\, \partial_{\tilde{\phi}}\ .
\end{eqnarray}
If differential operators are operated to scalar fields  
$\Phi$ in a form of Eq.(\ref{separation001}), 
the invariant relation becomes for fixed $\theta$ and  
$r=r_{+}$ (on the horizon) as 
\begin{eqnarray}
-\omega t+m\phi= -\tilde{\omega}\tilde{t}+\tilde{m}\tilde{\phi}\ ,
\end{eqnarray}
where the co-rotating coordinates, 
frequency and azimuthal angular momentum are     
\begin{eqnarray}
\tilde{t}&=&N_{\theta}
\left(t-\frac{a\sin^2{\theta}}{\Xi}\phi\right)\, ,\ 
\tilde{\phi}=N_{\theta}
\left(\phi-\frac{a\Xi}{r_{+}^2+a^2}t\right)\ , \nonumber\\
\tilde{\omega}&=&N_{\theta}
(\omega-\Omega_{\rm H}m)\ , \ 
\tilde{m}=N_{\theta}
\left(m-\frac{a\sin^2{\theta}}{\Xi}\omega\right)
\ , \label{co-rot-inf-rel}
\end{eqnarray}
with the normalization factor 
$N_{\theta}=\sqrt{(r_{+}^2+a^2)/(r_{+}^2+a^2\cos^2{\theta})}$.
The relations between 
the co-rotating frame observer at the black hole horizon 
\footnote{This observer is called as 
{\it zero angular momentum observer} (ZAMO).}
and the infinite frame observer in Eq.(\ref{co-rot-inf-rel}) 
leads the positive value of $\rm{Re}\,\omega-\Omega_{\rm H}\it{m}$:  
\begin{eqnarray} 
0<\rm{Re}\,\tilde{\omega} \ \ \ \Rightarrow 
\ \ \ 0<\rm{Re}\,{\omega}-\Omega_{\rm H} \it{m},
\label{real-part}
\end{eqnarray}
and the negative value of ${\rm Im}\,\omega$:
\begin{eqnarray}
\rm{Im}\,\tilde{\omega}<0 \ \ \ \Rightarrow 
\ \ \ \rm{Im}\,{\omega}<0\ .
\label{imaginary-part}
\end{eqnarray}
This result suggests that the unstable super-radiant modes 
do not occur. 
It is worthwhile to note 
that the stable type of super-radiant modes can occur for the negative values
of frequency ${\rm Re}\,\omega<0$ under the condition of Eq.(\ref{real-part}).


\begin{thebibliography}{99}
\bibitem{eckart001}
A. Eckart and R. Genzel, Nature {\bf 383} (1996) 415.
\bibitem{herrnstein001}
J.R. Harrnstein et al., Nature {\bf 400} (1999) 539. 
\bibitem{MBH003}
R. Sch\"{o}del, et al., Nature {\bf 419} (2002) 694.
\bibitem{exact001}
H. Stephani, D. Kramer, M.A.H. Maccallum, 
C. Hoenselaers and E. Herlt, 
{\it Exact Solutions to Einstein's Field Equations Second Edition}, 
Cambridge University Press, London (2003).
\bibitem{WMAP001}
D.N. Spergel et al., 
Astrophys. J. Suppl., 148 (2003) 175. 
\bibitem{maldacena001} 
J. Maldacena, Adv. Theor. Math. Phys., {\bf 2} (1998) 231.
\bibitem{myers001}
R.C. Myers and M.J.Perry, Ann. Phys. (Berlin) {\bf 172} (1986) 304.
\bibitem{gibbons001}
G.W. Gibbons, H. Lu, D.N. Page and C.N. Pope, J. Geom. Phys. 
{\bf 54} (2004) 49; 
G.W. Gibbons, M.J. Perry and C.N. Pope, Class. Quant. Grav.  
{\bf 22} (2005) 1503.
 \bibitem{chen001}
W. Chen, H. Lu and C.N. Pope, 
Class. Quant. Grav. {\bf 23} (2006) 5323.
\bibitem{green001}
M.B. Green, J.H. Schwarz and E. Witten, 
{\it Superstring theory}, Cambridge University Press (1987).
\bibitem{polchinski001}
J. Polchinski, {\it String Theory}, volume I and II, Cambridge 
University Press, London (1998).
\bibitem{randall001}
L. Randall and R. Sandrum, Phys. Rev. Lett. {\bf 86} (1999) 3370; 
{\bf 83} (1999) 4690. 
\bibitem{kaku001}
See, for example, 
M. Kaku, {\it String, Conformal Fields, and M-Theory}, 
Springer, New York (1999).    
\bibitem{birrell001}
N.D. Birrell and P.C.W. Davies, 
{\it Quantum Fields in Curved Space}, Cambridge University Press, 
London (1982).
\bibitem{chandra001}
S. Chandrasekhar, {\it The Mathematical Theory of Black Holes}, 
Oxford University Press, New York (1983).
\bibitem{kokkotas001}
K.D. Kokkotas and B.G. Schmidt, {\it Living Reviews in Relativity}, 
Vol. {\bf 2}, 
published by the Max Plnck Institute for Gravitational Physics, 
Albert Einstein Institute, Gernmany (1999). 
\bibitem{teukolsky000}
S.A. Teukolsky, Astrophys. J. {\bf 185} (1973) 635.
\bibitem{leaver001}
For a review article, see for example, 
E.W. Leaver, J. math. Phys. {\bf 27} (1986) 1238. 
\bibitem{takasugi001}
Analytical solutions of Kerr-de Sitter black holes  
by use of the series of hyper-geometric functions are given by   
H. Suzuki, E. Takasugi and H. Umetsu,
Prog. Theor. Phys. {\bf 100} (1998) 491.
\bibitem{heun001}
For a summary book, see for example, 
{\it Heun's Differential Equations}, edited by 
A. Ronveaux, Oxford University Press, Oxfird New York, (1995).

\bibitem{kodama001}
For review articles, for example, 
H. Kodama, J. Korean Phys. Soc. {\bf 45} (2004) S68; 
H. Kodama, {\it Perturbations and Stability of 
Higher-Dimensional Black Holes}, arXivw:0712.2703; 
H. Kodama, Prog. Theor. Phys. Supplement {\bf 172} (2008) 11.

\bibitem{detweiler001}
S. Detweiler, Phys. Rev. {\bf D22} (1980) 2323.
\bibitem{nambu001} 
H. Furusawa and Y. Nambu,  
Prog. Theor. Phys. {\bf 112} (2004) 983. 
\bibitem{dotti001}
G. Dotti and R.J. Gleiser, Class. Quantum Grav., 
{\bf 25} (2008)
245012;arXiv:0805.4306.
\bibitem{hawking000}
S.W. Hawking and H.S. Reall, Phys. Rev. {\bf D61} (1999) 024014.

\bibitem{teukolsky001}
J.M. Bardeen, W.H. Press and S.A. Teukolsky, 
Astrophys. J. {\bf 178} (1972) 347;
S.A. Teukolsky and W.H. Press,  
Astrophys. J. {\bf 193} (1974) 443. 
\bibitem{cardoso001}
V. Cardoso and O.J.C. Dais, Phys. Rev. {\bf D70} (2004) 084011.
\bibitem{cardoso002} 
V. Cardoso, {\'O}.J.C. Dias, J.P.S. Lemos and S. Yoshida, 
Phys.Rev. {\bf D70} (2004) 044039. 
\bibitem{press001}
W.H. Press and S.A. Teukolsky, 
Mature {\bf 238} (1972) 211. 
\bibitem{kunduri001}
H.K. Kunduri, J. Lucoetti and H.S. Real, 
Phys. Rev. {\bf D74} (2006) 084021.  

\bibitem{bekenstein001}
J.D. Bekenstein, Phys. Rev. {\bf D7} (1973) 2333.
\bibitem{bardeen001}
M. Bardeen, B. Carter and S.W. Hawking, Comm. Math. Phys. 
{\bf 31} (1973) 161. 
\bibitem{hawking001}   
S.W. Hawking, Comm. Math. Phys.{\bf 43} (1975) 199.
\bibitem{carlip001}
S. Carlip, {\it Black Hole Thermodynamics and Statistical Mechanics}, 
arXiv:0807.4520.
\bibitem{strominger001}
A. Strominger and C. Vafa, Phys. Lett. {\bf B379} (1996) 99.
\bibitem{thooft001}
G. 't Hooft, Nucl. Phys. {\bf B256} (1985) 727.
\bibitem{mukohyama001}
S. Mukohyama, Phys. Rev. {\bf D61} (2000) 124021.
\bibitem{mukohyama002}
S. Mukohyama and W. Israel, Phys. Rev. {\bf D58} (1998) 104005.
\bibitem{btz001}
M. Ba{\~{n}}ados, C. Teitelboim and J. Zanelli, 
Phys. Rev. Lett. {\bf 69} (1992) 1849. 
\bibitem{ichinose001}
I. Ichinose and Y. Satoh, Nucl. Phys. {\bf 447} (1995) 340. 
\bibitem{swkim001}
S.-W. Kim, W.T. Kim, Y.-J. Park and H. Shin, 
Phys. Lett. {\bf B392} (1997) 311.
\bibitem{fatibene001}
L. Fatibene, M. Ferraris, M. Fracaviglia and M. Raiteri, 
Phys. Rev. {\bf D60} (1999) 124012.
\bibitem{ho001}
J. Ho and G. Kang, Phys. Lett. {\bf B445} (1998) 27. 
 
\bibitem{kenmoku003}
M. Kenmoku, M. Kuwata and K. Shigemoto, Class. Quantum Grav. 
{\bf 25} (2008) 145016; arXiv:0801.2044.
\bibitem{kuwata001}
M. Kuwata, M. Kenmoku and K. Shigemoto, Prog. Theor. Phys. 
{\bf 119} (2008) 939; arXiv:0803.0604.  
\bibitem{birmingham001} 
D. Birmingham, Phys.Rev. {\bf D64} (2001) 064024. 

\bibitem{carter001}
B. Caeter, Commun. Math. Phys. {\bf 10} (1968) 280.
\bibitem{boyer001}
R.H. Boyer and R.W. Lindquist, J. Math. Phys. 
(1967) 265.
\bibitem{price001}
R.H. Price and V. Husain, Phys. Rev. Lett., {\bf 30} (1992) 1973. 
\bibitem{weinberg001}
See, for example, S. Weinberg,
{\it The Quantum Theory of fields}, Vol. {\bf I}, 
Cambridge University Press, Cambridge (1995).

\bibitem{hawking002}
J.B. Hartle and S.W. Hawking, 
Phys.Rev. {\bf D13} (1976) 2188. 

\bibitem{kenmoku001}
M. Kenmoku, K. Ishimoto, K.K. Nandi and K. Shigemoto, 
Phys.Rev. {\bf D73} (2006) 064004. 
\bibitem{kenmoku002}
M. Kenmoku and Y. Kobayashi,  
Class.Quant.Grav. {\bf 23} (2006) 6257-6274.
\bibitem{kenmoku005}
M. Kenmoku, in preparetion.  
\bibitem{kenmoku004}
M. Kenmoku, Y. Kobayashi, H. Terao and K. Shigemoto, 
in preparation.

\end{thebibliography}
\end{document}